\documentclass[12pt,a4paper,twoside]{article}

\usepackage{epsfig}
\usepackage{baltlat}
\begin{document}
\ \ 
\vspace{0.5mm}

\setcounter{page}{1}
\vspace{8mm}

\titlehead{Baltic Astronomy, vol.11, ***--***, 2002.}

\titleb{WHITEDWARF.ORG - ESTABLISHING A PERMANENT ENDOWMENT FOR THE 
WHOLE EARTH TELESCOPE}

\begin{authorl}
\authorb{Travis S.~Metcalfe}{1}
\end{authorl}

\begin{addressl}
\addressb{1}{White Dwarf Research Corporation, Austin, Texas U.S.A.}
\end{addressl}

\submitb{Received August 8, 2002}

\begin{abstract}
White Dwarf Research Corporation is a 501(c)(3) non-profit
organization dedicated to scientific research and public education on
topics relevant to white dwarf stars. It was founded in 1999 in Austin,
Texas to help fulfill the need for an alternative research center where
scarce funding dollars could be used more efficiently, and to provide a
direct link between astronomers who study white dwarf stars and the
general public. Due to its administrative simplicity, WDRC can facilitate
the funding of multi-institutional and international collaborations,
provide seamless grant portability, minimize overhead rates, and actively
seek non-governmental funding sources. I describe the motivation for, and
current status of, one of the long-term goals of WDRC: to establish a
permanent endowment for the operation of the Whole Earth Telescope. I pay
particular attention to fund-raising efforts through the website at {\tt
http://WhiteDwarf.org/donate/} 
\end{abstract}


\resthead{WhiteDwarf.org}{T.S.~Metcalfe}

\sectionb{1}{INTRODUCTION}

Several years ago, while I was a graduate student at the University of
Texas in Austin, I recognized the need for an alternative to traditional
academic and government research centers where the overhead rates, or
``indirect costs'' negotiated with federal funding agencies, are consuming
a growing slice of research budgets. This need seemed particularly urgent
since astronomical research funding has actually been shrinking when
adjusted for inflation, and many institutions appear to be providing fewer
services even while their overhead rates increase. Furthermore, many of
the services they {\it are} providing---fast computers, Internet access,
and libraries---can now be purchased at better prices through consumer
markets. I soon learned that I was not the first to make this realization.
An early pioneer was Research Corporation (founded in 1912), but the basic
idea has also been applied more recently by Eureka Scientific (in 1992),
Extrasolar Research Corporation (in 1996), and probably many others that I
do not know about.

\sectionb{2}{HISTORY}

Of course, recognizing the {\it need} for an alternative does not
automatically lead a poor graduate student with no legal experience to
{\it establish} an alternative. The idea of White Dwarf Research
Corporation would never have become a reality had it not been for two
other factors:  (1) creating a non-profit corporation in Texas is simple,
and (2) it is also inexpensive.  The concept of a corporation historically
involved the granting of limited liability by kings or governments to an
artificial legal entity created, for specific projects that were in the
public interest, with finite lifetimes. But over the years the scope and
definition of the corporation has slowly expanded (Derber 1998).  In the
state of Texas, anyone can create a non-profit organization with no
personal liability, a perpetual existence, and no specific public interest
requirement---all for the low price of \$25. The office of the Secretary
of State in Texas, and the Internal Revenue Service even offer samples of
the required incorporation documents to make it easy to satisfy the
(minimal) requirements, and to help ensure that the organization will be
tax-exempt for specific purposes (e.g. scientific research). After a few
nights of reading at the campus Law library, I had everything I needed to
start a corporation.

The most important feature of a non-profit organization, from a
fund-raising standpoint, is the recognition of tax-exempt status from the
IRS. This allows benefactors to deduct the full amount of their donations
from the incomes they report to the federal government, which can
substantially reduce their tax liability. If the financial structure of
the non-profit organization is relatively simple, then the application
form that needs to be filed with the IRS to obtain tax-exempt status is
not too difficult. If the organization anticipates an average of less than
\$10,000 of revenue per year during the first four years of operation, then
the filing fee for this document is only \$150. Otherwise, it is \$500,
which is by far the single most expensive part of the process. Within 6
months of filing the application, I received a letter from the IRS
recognizing the tax-exempt status of White Dwarf Research Corporation. I
was ready to begin raising money for research and public outreach.

The first grant to WDRC came from the Fund for Astrophysical Research, a
private foundation that makes small awards to support specific projects.
In December 2000, less than a year after receiving tax-exempt status, WDRC
received a check in the mail for \$2,812 to support the construction of a
small parallel computer cluster. We used this grant to attract matching
funds from another program through the American Astronomical Society, and
obtained an additional \$4,288 for the project in July 2001. We purchased
the parts for the computer facility a few months later, and it has already
contributed to a published research paper (see Metcalfe et al.~2002).

There have been several small private contributions to WDRC, but they
have all come from acquaintances of the members of the board of directors.
This is fine, but it would be more comforting to attract donations from
members of the general public, as a sign of trust and support for basic
research. To facilitate this goal, the most recent addition to the website
at {\tt WhiteDwarf.org} is a page describing specific projects that are
seeking funds (including the Whole Earth Telescope project), and a secure
online donation process. This service is provided free of charge by the
Network for Good, a non-profit organization that attempts to guide
potential donors through a database of non-profit organizations. It is now
possible, with the click of a button, for interested donors to use their
credit card to make a contribution to WDRC. The full amount of the
donation is then transferred to WDRC's bank account, and a notice is sent
indicating whether it is earmarked for a specific project or purpose.

The alternative funding machinery is now fully in place: we have
established a non-profit organization, it has tax-exempt status, it has
successfully received and allocated funding for a small research project,
and the website has the capability for secure online donation. What can we
do with this new tool, and why should we bother?

\sectionb{3}{MOTIVATION}

The Whole Earth Telescope (WET; Nather et al.~1990) operates like any
large research facility: it produces significant results on time-scales
that can be considerably longer than than the 3-year funding cycles of
federal granting agencies. It is natural for these agencies to request a
specific and detailed accounting of the results that arise from their
investment in a research project. But this can sometimes lead to a
conflict between the political needs of these institutions and the
scientific needs of the project. In the short-term, we can adapt the way
the Whole Earth Telescope operates to respond to these political
pressures, but in the long-term it may be better to find an alternative
source of funding to alleviate the pressures altogether. I believe that
White Dwarf Research Corporation can offer this alternative, by providing
an administrative vehicle for the establishment of a permanent endowment
to operate the Whole Earth Telescope. If the WET collaboration could
identify a group of donors who can collectively make contributions
totaling \$1-2 million, WDRC could easily provide the \$50-100 thousand
annual operating budget of the WET through secure investments that never
diminish the endowment. The WET could free itself from the tyranny of the
government funding cycle.

For an individual, \$1 million is a lot of money. But there are many ways
to reach this funding goal through a larger group of people. We could make
an appeal to 100,000 people to contribute only \$10, or just 10,000 people
to contribute \$100. This is comparable to what people might spend on a
magazine subscription, or to support their favorite political candidate.

For most transnational corporations, \$1 million is a tiny fraction of a
typical annual advertising budget. It is realistic to think that we may be
able to identify ten \$100,000 donors, or one hundred \$10,000 donors.
Some people may think that no transnational corporation could possibly be
{\it interested} in helping to fund the WET. What could possibly be in it
for them, besides a warm fuzzy feeling inside?

During a WET campaign in 1998 on the massive DAV star BPM~37093, a test of
the theory of stellar crystallization, a flurry of reports surfaced in the
popular press about this ``Diamond in the Sky''. According to Steve
Kawaler, a large diamond company saw these press reports and contacted the
headquarters in Iowa during the run to ask what it would cost to sponsor
the run on this object. When he replied that the price tag would be in the
neighborhood of \$100,000 they replied, ``Oh, is that all?''. The point is
that if we can be creative and capture the public imagination, and if we
are receptive to seemingly unlikely sponsors, we may be able to get more
science done with less hassle.

The right approach is probably to seek some mixture of these funding
possibilities. The numbers are not staggering: with some collective effort
now, the goal is realistic and will yield a payoff far into the future. In
the worst case, we might fail to reach the full funding level, but the
interest on the partial sum could still provide flexible supplemental
funding to the WET---perhaps enough to send the PI to headquarters, a
graduate student to the observatory, or to purchase GPS units for
distribution around the globe.

\sectionb{4}{CURRENT STATUS}

The purpose of this paper is not to ``pass around a hat'' and beg a group 
of underpaid scientists for money. The point is to make it clear that
together we can come up with more creative ways of funding our
collaboration, and the pipeline for incoming funds already exists: White
Dwarf Research Corporation. Now that you know about it, do what you can to
raise awareness about it. Be receptive to any offers of support, and
direct potential donors to {\tt http://WhiteDwarf.org/donate/} or to the
Whole Earth Telescope web page. Do your best to think about ways to make
our work appealing, or better yet {\it fascinating} to the general public,
and imagine which private companies might want to help fund it.

For my part, I have started the Whole Earth Telescope endowment with a
donation of \$50 through the website (just to make sure it really does
work as advertised---which it does). Within the next three years, before
the current NSF grant expires, let's see if we can raise enough money to
operate the WET for as long as the white dwarf stars shine. 
\vskip7mm

ACKNOWLEDGMENTS.\ I would like to thank Ed Nather and Don Winget for 
their early participation in WDRC, and Jim Liebert, Mike Montgomery
and Hugh Van Horn for agreeing to serve as members of the Board of 
Directors.
\goodbreak

\References

\ref{Derber, C. 1998, Corporation Nation: how corporations are taking over 
our lives and what we can do about it (New York: St.~Martin's Press)}

\ref{Metcalfe, T. S., Salaris, M., \& Winget, D. E. 2002, ApJ, 573, 803}

\ref{Nather, R. E. et al.~1990, ApJ, 361, 309}

\end{document}